\documentclass[12pt]{article}

\usepackage{amsmath,amssymb,latexsym,theorem,graphicx,amscd}
\usepackage[all]{xy}
\CompileMatrices

\usepackage{color}
\definecolor{dblue}{rgb}{0,0,.7}
\definecolor{indigo}{RGB}{50,0,105}

\usepackage[hyperindex=true, colorlinks=true, pagecolor=dblue, bookmarksopen=false, urlcolor=dblue, linktocpage,linkcolor=dblue, citecolor=indigo]{hyperref}
\def\P{{\mathbb P}} \def\O{\mathcal{O}} \def\C{{\mathbb C}}\def\ff#1#2{{\textstyle\dfrac{#1}{#2}}}
\newcommand{\be}{\begin{equation}}
\newcommand{\ee}{\end{equation}}
\newcommand{\eq}[1]{Eq. (\ref{eq:#1})}
\newcommand{\Aphys}{{A^{(phys)}}}
\newcommand{\bAphys}{{{\bar A}^{(phys)}}}
\newcommand{\Amath}{{A^{(math)}}}
\newcommand{\bAmath}{{{\bar A}^{(math)}}}
\newcommand{\bbc}{{\mathbb C}}

\newcommand{\Aut}{{\operatorname{Aut}}}
\newcommand{\vol}{{\text{vol} }}
\newcommand{\Tr}{{\text{ Tr}~ }}

\newcommand{\End}{{\text{ End}~ }}

\newcommand\ba{\bar{a}}
\newcommand\bb{{\bar{b}}}

\newcommand\bZ{\bar{Z}}

\newcommand\bD{\bar{D}}
\newcommand\bA{\bar{A}}

\newcommand\bbeta{{\bar{\beta}}}
\newcommand\bOmega{\bar{\Omega}}
\newcommand\bpartial{\bar{\partial}}
\newcommand\bj{{\bar{j}}}
\newcommand\bi{{\bar{i}}}

\newcommand\bs{{\bar{s}}}

\newcommand{\vev}[1]{\langle{#1}\rangle}
\newcommand{\cp}[1]{{\mathbb P}^{#1}}

\newcommand{\om}{{\mathcal O}_M}

\newcommand{\cL}{{\mathcal L}}

\newcommand{\p}[1]{{\mathbb P}(#1)}

\newtheorem{theo}{Theorem}[section]

\newtheorem{prop}[theo]{Proposition}

\newtheorem{conj}[theo]{Conjecture}
{\theorembodyfont{\rmfamily} }
{\theorembodyfont{\rmfamily} }

\def\pplogo{\vbox{\kern-\headheight\kern -29pt
\halign{##&##\hfil\cr&{\ppnumber}\cr\rule{0pt}{2.5ex}&\ppdate\cr}}}
\makeatletter
\def\ps@firstpage{\ps@empty \def\@oddhead{\hss\pplogo}%
  \let\@evenhead\@oddhead 
}
\def\maketitle{\par
 \begingroup
 \def\thefootnote{\fnsymbol{footnote}}
 \def\@makefnmark{\hbox{$^{\@thefnmark}$\hss}}
 \if@twocolumn
 \twocolumn[\@maketitle]
 \else \newpage
 \global\@topnum\z@ \@maketitle \fi\thispagestyle{firstpage}\@thanks
 \endgroup
 \setcounter{footnote}{0}
 \let\maketitle\relax
 \let\@maketitle\relax
 \gdef\@thanks{}\gdef\@author{}\gdef\@title{}\let\thanks\relax}
\makeatother

\begin{document}
\setcounter{page}0
\def\ppnumber{\vbox{\baselineskip14pt
\hbox{hep-th/0606261}}}
\def\ppdate{RUNHETC-06-15} \date{}

\title{\bf \LARGE 
	Numerical solution to the hermitian Yang-Mills equation on the Fermat quintic	\\[10mm]}
\author{\small
     {{\bf Michael R. Douglas}$\stackrel{\&}{\bf ,}$
     {\bf Robert L.~Karp}, 
     {\bf Sergio Lukic}
     {\bf and Ren\'e Reinbacher}} \\	
[10mm]
\normalsize  Department of Physics, Rutgers University \\
\normalsize Piscataway, NJ 08854-8019  USA			\\	[10mm]
\normalsize $^\&$ I.H.E.S.,  Le Bois-Marie, Bures-sur-Yvette, 91440 France}

{\hfuzz=10cm\maketitle}

\vskip 1cm

\begin{abstract}
\normalsize
\noindent
We develop an iterative method for finding solutions
to the hermitian Yang-Mills equation on stable holomorphic vector
bundles, following ideas recently developed by Donaldson. 
As illustrations, we construct numerically the hermitian Einstein metrics
on the tangent bundle and a rank three vector bundle on
$\mathbb{P}^2$. In addition, we find a hermitian Yang-Mills connection
on a stable rank three vector bundle on the Fermat quintic.
\end{abstract}

\vfil\break

\tableofcontents

\section{Introduction}    \label{s:intro}

The modern study of compactification of higher dimensional theories
can be divided into two general branches.  The first makes use of
compactification manifolds with a good deal of symmetry, such as the
torus, sphere, squashed spheres and so on.  Such spaces have
explicitly known metrics, allowing explicit solutions of the equations
of motion, and explicit Kaluza-Klein reduction.  These solutions have many
applications, such as supergravity duals of large $N$ gauge theories;
however their high degree of symmetry tends to be a problem in trying to
obtain models with the level of complexity of the Standard Model or its
often-postulated extensions.

The second branch makes use of manifolds for which the relevant metrics
are known to exist by general theorems, but for which explicit
expressions are not known.  The most famous examples are the
Ricci-flat Kahler metrics conjectured to exist by Calabi and proven
to exist by Yau \cite{Yau}.
In 1985, it was proposed by Candelas {\it et al}  \cite{CHSW}
that compactification of the heterotic string on a Calabi-Yau manifold
could lead to quasi-realistic theories of particle physics, containing
grand unified extensions of the Standard Model and low energy
supersymmetry.
Since then, other metrics of this type, such as $G_2$
holonomy metrics, have been used in quasi-realistic compactifications;
see for example \cite{Acharya:Witten}.

Over the subsequent years, many tricks were developed to bypass the
difficulties posed by not knowing the compactification metric.  These
tricks began with the algebraic geometry behind the theorems of Yau
and Donaldson-Uhlenbeck-Yau, and gradually evolved into entire
branches of mathematical physics, such as topological string theory and
special geometry.  To drastically oversimplify, the
general picture is that certain ``protected'' quantities in the four
dimensional Effective Field Theory (EFT), such as the superpotential
in theories with four supercharges, and the prepotential in theories
with eight supercharges, can be computed using techniques combining
algebraic geometry with physical ideas.  Other quantities, such as the
Kahler potential in theories with four supercharges, cannot be
computed directly.  Since a good deal of important physics depends on
the Kahler potential -- precise values of particle masses, and the
existence and stability of supersymmetry breaking vacua, this
situation is not very satisfactory.

Almost all present knowledge about the Kahler potential in
the EFT comes from studying expansions around more computable limits.
The best known example is the case of $N=1$ compactifications which
contain $N=2$ subsectors, such as heterotic $(2,2)$ models, or type II
on Calabi-Yau orientifolds.  In these cases, there is a limit in which
part of the $N=1$ Kahler potential becomes equal to that of the
related $N=2$ theory, which is computable using special geometry.
Other examples include the solvable orbifold or Gepner model limits,
at which the entire Kahler potential is computable in principle
using CFT techniques.  However, it is not clear at present how
representative such results are of the general case.  Even a limited
ability to compute in the general case would allow studying this
question.

One completely general technique for addressing such problems is to
compute the Ricci-flat metrics and related quantities numerically.
Numerical methods are unavoidable in other areas of physics, beginning
with such seemingly elementary problems as computing the
spectrum of the helium atom or integrating Newton's equations for
the three body problem in
celestial mechanics; it would be surprising if string theory 
could avoid this.  To bring string theory closer to a possible
confrontation with real data, for example from collider physics, it
may be valuable to develop these missing parts of the theory of
compactification.

In this work, we make a start in this direction by showing two 
things. First, we review how to use existing mathematical techniques
to numerically approximate metrics on Kahler manifolds, along lines
recently developed by Donaldson \cite{Donaldson:numeric}.  Second, we
extend these mathematical techniques to hermitian Yang-Mills
connections.  
It will be clear that these techniques could
be pushed to compute higher order terms, metrics on moduli spaces, and
the like.  A subsequent paper will explain the numerical methods in
more detail and do some simple computations of terms in the
EFT for compactification on a
quintic Calabi-Yau 3-manifold.

Our direct inspirations are Donaldson's work \cite{Donaldson:numeric}
on numerical approximation of metrics, and of Wang
\cite{Wang:metricsbundles} developing the corresponding mathematics
for vector bundles.  We can also cite Headrick and Wiseman
\cite{Headrick:Wiseman}, who made a pioneering numerical study of the
K3 metric using position-space methods.  Finally, the first author is 
particularly indebted to Bernie Shiffman and Steve Zelditch for teaching him
the basics of asymptotic analysis on holomorphic line bundles, and for
advice in the early stages of this project, in particular for
pointing out Wang's work.

Let us briefly explain the problem and survey some of the approaches
one might take towards it, before beginning the detailed development
in section 2.  Following \cite{CHSW}, the derivation of the matter
Lagrangian in a heterotic compactification on a Calabi-Yau $X$ 
carrying a bundle $V$ involves the following steps:
\begin{enumerate}
\item Find the Ricci-flat metric $g_{ij}$ (with specified moduli) on $X$.
\item Find the hermitian Yang-Mills connection $A_i$ on $V$.
\item Find the zero modes $\psi^\alpha$
of the Dirac operator.  As is standard, on a Kahler manifold this
amounts to finding harmonic differential forms $\psi$ valued in $V$,
{\it i.e.} solutions of 
$0=(\bpartial+\bA)\psi=(\bpartial+\bA)^*\psi$,
where $*$ denotes the adjoint operator.  
\item Find an orthonormal basis of forms $\psi$.
\item Compute the integrals over $X$ of wedge products of these forms
to get the superpotential.
\end{enumerate}
The key step for us is (4).  Existing methods for computing the
superpotential, such as \cite{GSW2,Candelas:1987rx,Donagi:2006yf}, accomplish step (5) without needing the
results of (1) and (2), by using unnormalized zero modes.  This leads
to a superpotential defined in terms of fields whose kinetic term is
obtained from ``some'' unknown Kahler potential.  To do better, we must
either derive normalized zero modes in (4) for use in (5), or else take
the zero modes used in (5) and compute their normalizations using the
explicit metric from (1).

There seems to be no way of doing this without some knowledge of the
Ricci-flat metric and thus the first step is to choose some
approximation scheme for this metric.  One's first thought might be to
follow standard practice in numerical relativity, as done in
\cite{Headrick:Wiseman}, and introduce a six dimensional lattice which
is a discrete approximation to the manifold $X$; in other words a
position space approach.  Taking the Kahler potential $K$ as the basic
dynamical variable, Einstein's equations reduce to the complex
Monge-Ampere equation \be\label{eq:mongeampere} \det(\partial\bpartial
K) = \Omega\wedge\bOmega \ee which can be solved by relaxation
methods.  One would then need to find similar lattice approximations
for the connection on $V$ and the zero modes.
 
An alternative approach, introduced by Donaldson
\cite{Donaldson:numeric}, is to use the natural embedding of $X$ into
$\P^{N-1}$ provided by the $N$ sections of an ample line bundle $L^k$ (we will
explain this in detail below).  We then take as a candidate
approximating metric on $X$ the pull-back of a Fubini-Study metric on
$\P^{N-1}$.  Such a metric is defined by an $N\times N$ hermitian
matrix. By suitably choosing this matrix we can try to make the
associated Fubini-Study metric restrict to $X$ in such a way that it
gives a good approximation to the Ricci-flat metric on $X$.

A major advantage of this approach is that it avoids the complications and
arbitrariness involved in choosing an explicit discretization of $X$;
rather the entire approximation scheme follows from a single parameter
$k$, the scale of the first Chern class of $L$.  Subsequent
mathematical development reveals more structure which can be used to our
advantage.  For example, a very natural approximation to the Ricci-flat
metric, which becomes exact as $k\rightarrow\infty$, is the so-called
``balanced'' metric.  In a sense, to be described below, this is
the metric for which the embedding of $X$ into $\P^{N-1}$ has its center
of mass at the ``origin''.  It also satisfies a simple fixed point condition
which can be used for relaxation, solving step (1).

Another advantage, which is key for the present application, is that
Donaldson's method can be naturally extended to study holomorphic
vector bundles on $X$.  There is a standard relation between
holomorphic connections and hermitian metrics, which we review in
section 2, in which step (2) of the above prescription is turned into
the problem of finding a hermitian-Einstein metric on a vector
bundle. For illustrative purposes we will explicitly study
hermitian-Einstein metrics on two spaces: complex projective space
$\P^n$ and the Fermat quintic threefold.

The organization of the paper is as follows. In Section~2 we provide
an overview of the geometric background needed for our construction,
in particular we will describe Donaldson's approach for getting
metrics of constant scalar curvature. In Section 3 we explain a
numerical approximation to the hermitian Einstein metric on a
holomorphic vector bundle by a simple adaptation of Donaldson's
scheme, building on mathematical work of Wang.  
In section 4 we focus
on several explicit examples. Here we describe some of our numerical
methods and results in detail. By design we are also able to test our
approximation scheme for $T\P^2(k)$, where $T\P^2$ is the holomorphic
tangent bundle of $\P^2$, since in this case one has an analytic
solution.

\section{Metrics of constant scalar curvature}

We follow the plan outlined in the introduction, beginning with step (1). Let
$X$ be an $n$-dimensional complex Kahler ``compactification manifold.''
Since we are not assuming it is a valid string theory background, we 
can generalize the discussion to arbitrary $n$ and first Chern class $c_1(X)$.

The basic example we have in mind is the complex projective space
$\cp{n}$, parameterized by the standard homogenous coordinates
$\{Z_i\}_{i=0}^n$, up to the identification $\{Z_i\} \cong \{\lambda
Z_i\}$ for $\lambda\in\bbc^*$.  The Kahler potential
\be\label{eq:unitFS} K_{FS} = k~ \log \left(\sum_{i=0}^n |Z_i|^2
\right) \ee defines the Fubini-Study metric on $\P^n$, with $SU(n+1)$
symmetry $Z_i\rightarrow g_i^j Z_j$, and $g\in U(n+1)$.  The parameter
$k$ controls the Kahler class $\omega=\partial\bpartial K_{FS}$.

Our other general example 
is the hypersurface defined by the vanishing of a degree $d$ polynomial 
in $\P^n$:
\be\label{eq:hyperQ}
f(Z) = \sum_{i_1\cdots i_d} c^{i_1\cdots i_d} Z_{i_1}\ldots Z_{i_d}.
\ee
For $n=4$ and $d=5$ we get a quintic threefold $Q$.  Its complex
structure is determined by the $126$ parameters $c^{i_1\cdots i_5}$,
modulo the action of $GL(5,\C)$ on the $Z_i$'s,
which leaves $101$ parameters.  The generic member of 
this family is smooth, and has $b^{1,1}=1$. Therefore the
Kahler class is determined by a single real number.  A simple one
parameter family of Kahler metrics on $Q$ is obtained by pulling back the
Fubini-Study metric on $\cp{4}$, or equivalently interpreting \eq{unitFS}
as a Kahler potential on $Q$.  Of course this will not be Ricci-flat.

\subsection{Approximating Ricci-flat metrics by projective embedding}

We now want to find a larger space of Kahler metrics in which
to find a better approximation to the Ricci flat metric.
One simple generalization of \eq{unitFS} can be obtained by choosing an
$(n+1)\times (n+1)$ hermitian matrix $h^{i\bj}$, and writing
\be\label{eq:hFS}
K_{h} = k \log \left(\sum_{i,\bj=0}^n h^{i\bj} Z_i \bZ_j \right).
\ee
Of course, by making a linear redefinition of coordinates, we could
turn this back into \eq{unitFS}, but doing so would modify
\eq{hyperQ}.  Rather, by fixing \eq{hyperQ}, this way we get an $(n+1)^2$-parameter family of Kahler potentials.

Another way to think about this definition is to make the linear
redefinition taking $h$ to the identity.  In this case, the parameters
we are varying to control the metric are the extra $25$ parameters in
\eq{hyperQ} determining a specific embedding of $Q$ into $\P^4$.
While all of the embeddings are equivalent under a $GL(5,\C)$ action, once
we use the metric, we break this to $U(5)$; thus the set of metrics
we can obtain this way is parameterized by a $GL(5,\C)/U(5)$ homogeneous space.

A simple generalization to get more parameters could be motivated by
noticing that \eq{unitFS} is also equal to
\begin{eqnarray}
K_{FS,k} &=& \log \left(\sum_{i=0}^n |Z_i|^2 \right)^k \\
 &=& \log \left(\sum_{i_1,\cdots,i_k=0}^n
 Z_{i_1}\cdots Z_{i_k}\bZ_{i_1}\cdots\bZ_{i_k} \right) 
\end{eqnarray}
and generalizing this to
\be\label{eq:Kh}
K_{h,k} = \log \left(\sum_{i_1,\cdots,i_k,\bj_1,\cdots,\bj_k=0}^n
 h^{i_1\cdots i_k\bj_1\cdots \bj_k}
 Z_{i_1}\cdots Z_{i_k}\bZ_{\bj_1}\cdots\bZ_{\bj_k}\right) ,
\ee
which can again be interpreted as a Kahler potential on $Q$.
In simple terms, we are using higher degree polynomials as basis functions.
Now we have an $(n+1)^{2k}$-parameter family of metrics, and by taking
$k$ large we can imagine finding an arbitrarily good approximating metric
within this class.  

One way to find the best approximation to the Ricci-flat metric on $Q$
would be to write \eq{mongeampere} directly in these variables.  Note that
the holomorphic $(n,0)$-form $\Omega$ is known explicitly. For example, in the  coordinate patch where $Z_0\neq 0$ we can choose the local coordinates $w_i=Z_i/Z_0$, in terms of which
$$
\Omega = \frac{dw_1\, dw_2\, dw_3}{\partial f/\partial w_4} ,
$$
and thus one can write the volume form for the Ricci flat metric
explicitly,
\be\label{eq:cyvol}
{d\vol}_X  = \Omega\wedge\bOmega
\ee
without solving any equations.  One might then substitute \eq{Kh} into
\eq{mongeampere}, evaluate this at a set of points $p_i$, and solve
the resulting system of nonlinear equations.  These are rather
complicated, however, and furthermore we have introduced arbitrariness
in the choice of the $p_i$.  Now this arbitrariness can have
its uses, for example we might use it to place more points in
regions of large curvature.  On the other hand, it means that the
results will not have simple mathematical or physical interpretations,
except in the limit in which the number of points is so large that we
can ignore the discretization.\footnote{ Or unless we can come up with
a construction in which some sort of physical objects at the points
$p_i$ enforce the equations.}  Before investing a lot of effort into
their study, we should try to improve on this point.

\subsection{Balanced metrics}

There is a pretty construction that goes back to \cite{Tian:metrics}
which provides a more natural approximating metric, and a numerical
scheme which is guaranteed to converge to it.

First, we can systematize the construction which led to \eq{Kh}, by
noting that the basis functions are products of degree
$k$ holomorphic times  degree
$k$ antiholomorphic monomials.  Let the number
of independent holomorphic degree $k$ monomials be $N+1$; this is
the binomial coefficient $\binom{n+k} {k}$ 
for $\cp{n}$, and we will give it for $Q$ later.

Let us phrase this construction in a way 
which can be used for an arbitrary manifold $X$.  
We choose a holomorphic line bundle $\cL$ over $X$, 
with $N$ global sections.  Denote a complete basis of these as
$s_\alpha$, where $1\le\alpha\le N$, and consider the map
$$
i_k\colon X\longrightarrow \cp{N-1}
 \qquad i_k(Z_0,\ldots,Z_n) = (s_1(Z),s_2(Z),\ldots,s_N(Z)).
$$
The geometric picture is that each point in our original manifold $X$
(parameterized by the $Z_i$)
corresponds to a point in $\bbc^{N}$ parameterized by the sections
$s_\alpha$.  Since choosing a different frame for $\cL$ would produce
an overall rescaling $s_\alpha\rightarrow \lambda s_\alpha$, the overall
scale is undetermined. Granting that 
$s_1(Z),s_2(Z),\ldots,s_N(Z)$ 
do not vanish simultaneously, this gives us a map to $\P^{N-1}$. 

The simplest example is to embed $\cp{1}$ using $\cL=\O_{\P^1}(k)$ 
into $\cp{k}$. In this case the map is
$$
i_k(Z_0,Z_1) =
 (Z_0^k,~ Z_0^{k-1}Z_1,~Z_0^{k-2}Z_1^2,~\ldots,~ Z_0 Z_1^{k-1},~Z_1^k).
$$

In general we want this map to be an
embedding, {\it i.e.} that distinct points map to distinct points with
non-vanishing Jacobian.  In general, we can appeal to the Kodaira embedding
theorem, which asserts that for positive $\cL$
this will be true for all $\cL^k$ for some
$k\ge k_0$.  For non-singular quintics, this is true for $\om(k)$ for
all $k\ge 1$.  As a point of language, the pair of a
manifold $X$ with a positive line bundle $\cL$ is referred to as 
a {\em polarized manifold} $(X,\cL)$; the condition that this construction
provides an embedding for some $k$ is that $\cL$ is {\it ample}.

Now, we consider our family \eq{Kh} of candidate K\"ahler
potentials, and rewrite them as
$$
K_{h} = \log \left(\sum_{\alpha,\bbeta} h^{\alpha\bbeta}
 s_\alpha \bs_\bbeta \right)
$$
or simply
\be\label{eq:Ks}
K_{h} \equiv \log ||s||^2_h
\ee
for short, where $s_\alpha$ plays the role of a degree $k$ monomial.
We now have an $N^2$-parameter family of K\"ahler potentials, and will
seek a good approximating metric in this family.
Just as before, this amounts to using the pull-back of a Fubini-Study
metric from $\cp{N-1}$ as our trial metric.

Mathematically, the simplest interpretation of \eq{Ks} is that it
defines a hermitian metric on the line bundle
$\cL = \om(k)$.  This is a sesquilinear map 
from $\bar\cL\otimes \cL$ to smooth functions $C^\infty(X)$, here defined by
$$
(s,s') = e^{-K_{h}}\cdot\bar s\cdot s' =
 \frac{\bar s\cdot s'}{\sum_{\alpha,\bbeta} h^{\alpha\bbeta}
 s_\alpha \bs_\bbeta} .
$$
The point is that a change of frame, which acts on our explicit sections
as $s_\alpha\rightarrow \lambda s_\alpha$,
cancels out of this expression.\footnote{A possibly more familiar
physics use of this is in $N=1$ supergravity: taking
$K\rightarrow -K$ and  $s\rightarrow W$, one gets the standard 
expression for the gravitino mass $e^K|W|^2$.  In an example such as
the flux superpotential, in which $W$ is a sum of various terms $s_\alpha$
with constant coefficients, \eq{Ks} also applies to give $K$.}

This metric allow us to define an inner product between the global sections:
\be\label{eq:defH1}
H_{\alpha\bbeta} = \vev{s_\beta|s_\alpha} = i\int_X 
\frac{s_{\alpha}\bs_{\bbeta}}{||s||^2_h} \,d{\vol_X}.
\end{equation} 
This is the ``physical'' inner product in a sense we will explain further
below.
Note that it depends on $h$ in a nonlinear way, since $h$ appears
in the denominator.

Here $d\vol_X$ is a volume form on $X$, which has to be chosen.
If $X$ is Calabi-Yau, it is simplest to use \eq{cyvol} to define $d\vol_X$.
If $X$ is not Calabi-Yau, the standard choice of $d\vol_X$ is
to take $d\vol_\omega=\omega^n/n!$, where $\omega$ is the 
Kahler metric derived from \eq{Ks}.  This depends on $h$ as well, so
the expression is even more non-linear in $h$.

Thus, given $h$ and a basis of global sections $s_\alpha$, we could
compute the matrix of inner products \eq{defH1}.
Once we have it, we 
could make a linear redefinition, say $\tilde s = H^{-1/2} s$, and
go to a basis of orthonormal sections where
\be\label{eq:onH}
H_{\alpha\bbeta} = {\delta}_{\alpha\bbeta} .
\ee

On the other hand, \eq{Ks} also implicitly
defines a notion of orthonormal basis locally in the bundle, in which
\be
h^{\alpha\bbeta} = {\delta}^{\alpha\bbeta} .
\ee
This is {\it a priori} different from \eq{onH};
indeed we can freely postulate it when we write \eq{Ks}.
However, if the two notions agree,
$$
H_{\alpha\bbeta} = (h^{-1})_{\alpha\bbeta} ,
$$
then we can go to a basis of sections in which
\be\label{eq:balanced}
H_{\alpha\bbeta} = h^{\alpha\bbeta}  = {\delta}_{\alpha\bbeta} .
\ee
In this case, 
the embedding of $X$ in $\mathbb{P}^{N-1}$ using these sections
is called {\it balanced}.
More generally, we call
a polarized manifold $(X,\cL^k)$ balanced 
if such an embedding exists.

An equivalent definition of 
the balanced embedding is arrived at if we consider the
function on $X$ defined as
\begin{equation}\label{eq:bk}
 \rho(\omega)(x)=
 \sum_{\alpha,\bbeta} (H^{-1})^{\alpha\bbeta} (s_\alpha(x),\bs_\bbeta(x))
\end{equation}
or equivalently
$$
\rho(\omega)(x)= \sum_{\alpha} ||s_\alpha(x)||^2
$$
where the second sum is taken over an orthonormal basis in which
$H={\delta_{\alpha\beta}}$.  $X$ is balanced precisely when
$\rho(\omega)(x)$ is the constant function.

Many theorems have been proven about balanced manifolds.  Let us first
recall the following theorem of Donaldson (Theorem 1 in
\cite{Donaldson1}):
\begin{theo}\label{thm:unique} 
Suppose the automorphism group $\Aut(X,\cL)$ is discrete. If
$(X,\cL^k)$ is balanced, then the choice of basis in $H^0(X,\cL^k)$
such that $i_k(\cL)$ is balanced is unique up to the action of
$U(N)\times \mathbb{R}^*$.
\end{theo}
The condition on $\Aut(X,\cL)$, i.e., there are no continuous
symmetries, is true for the quintic $Q$.  This theorem then tells us
that, if a metric $h$ exists which gives a balanced embedding, it is
unique up to scale.

Given a balanced embedding, one defines the {\it balanced metric} on
$X$ as the pullback of the Fubini-Study metric (\ref{eq:Ks}):
\begin{equation}
\label{ }
\omega_k=\frac{2\pi}{k}i_k^*(\omega_{FS}),
\end{equation}
The cohomology class of the Kahler form $[\omega_k]=2\pi c_1(\cL)\in
H^2(X)$ is independent of $k$. Using these definitions Donaldson
proves that (Theorem 2 in \cite{Donaldson1}):
\begin{theo}\label{theo2}
Suppose $\Aut(X,\cL)$ is discrete and $(X,\cL^k)$ is balanced for
sufficiently large $k$. If the metrics $\omega_k$ converge
in the $C^\infty$ norm to some limit $\omega_\infty$ as $k\to \infty$, then
$\omega_\infty$ is a Kahler metric in the class
$2\pi c_1(\cL)$ with constant scalar curvature.
\end{theo}
The constant value of the scalar curvature is determined by $c_1(X)$,
and in particular for $c_1(X)=0$ the scalar curvature is zero.  Thus,
the balanced metrics $\omega_k$, in the large $k$ limit, converge to
the Ricci flat metric.

Therefore, if we can find the unique balanced metric for a given
$\cL$, it is a good candidate for approximating the Ricci flat metric
on $X$.  One may ask where the complex structure and Kahler moduli on
which this Ricci flat metric depends, are put in.  The complex
structure enters implicitly through the basis for holomorphic sections
$s_\alpha$, as we will see in examples below.  As for the Kahler
class, recall that this is determined, up to scale, to be $2\pi
c_1(\cL)$.  Of course, the Ricci flatness condition is scale
invariant, so the overall scale is irrelevant; however the point of
this is that if $b^{1,1}>1$, then by appropriately choosing $\cL$ we
choose a particular ray in the Kahler cone.  This will not be relevant
for our examples here but shows that in principle any Ricci-flat
Kahler metric could be approximated in this way.

\subsection{Finding the balanced metric}

In \cite{Donaldson2,Donaldson:numeric} Donaldson proposes a method to
determine the hermitian metric $h$ in \eq{Ks}, which will lead to a
balanced metric.  He defines the ``T operator'', which given a metric
$h$ computes the matrix $H$: 
\be\label{eq:defH} H_{\alpha\bbeta} =
T(h)_{\alpha\bbeta} \equiv \frac{N}{\vol(X)}\int_X \frac{s_\alpha
\bs_\bbeta}{||s||^2_h}\, {d\vol_X}
\end{equation} 
Now, suppose we find a fixed point of this operator,
$$
T(h) = h .
$$
Then, by a $GL(N)$ change of basis $s\rightarrow h^{-1/2}s$,
we can bring $h$ to the unit matrix,
which will produce the balanced embedding.

The simplest way to find a fixed point of an operator is to iterate
it.  If the operator is contracting, this is guaranteed to work.  In
our case we have the following theorem \cite{Donaldson1,Sano}:
\begin{theo}
Suppose that $\Aut(X,L)$ is discrete. If a balanced embedding
exists then, for any initial $G_0$ hermitian metric, the sequence $T^r(G_0)$
converges to the balanced metric $G$ as $r\to \infty$.
\end{theo}
Thus the $T$ operator can be used to find
approximate Ricci-flat metrics on Calabi-Yau manifolds, and more
generally approximate
constant scalar curvature Kahler metrics.  
In \cite{Donaldson:numeric} Donaldson
studies numerically explicit $\cp{1}$ and $K3$ examples.  We will discuss some
additional examples below.

\subsection{Balanced metrics and constant scalar curvature}\label{s:prs}

In this subsection we outline the reason why the limit of a family of
balanced metrics has constant scalar curvature. This is the content of
Theorem~\ref{theo2}. This will be very useful later on, when we
generalize the T-operator to vector bundles.

Note that the  function $\rho(\omega)$ is independent of the choice of orthonormal basis, and remains
unchanged if we replace $h$ by a constant scalar multiple. Therefore,
it is an invariant of the Kahler form. As discussed before, the balanced condition for
$(X,L^k)$ is equivalent to the existence of a metric $\omega_k$ such
that $\rho(\omega_k)$ is a constant function on $X$. The asymptotic behavior of
the ``density of states" function $\rho(\omega_k)$ as $k \to \infty$
for fixed $\omega$ has been studied in \cite{Tian:metrics,Catlin,Zelditch:Szego,LuZhiqin}. Note that for any metric
\begin{equation}
\label{eq-rho}
\int_X \rho_k(\omega) = N=\dim H^0(X,L^k)=a_0k^n +a_1k^{n-1}+\cdots,
\end{equation}
where the  coefficients $a_i$ can be determined using the Riemann-Roch formula.  Note that $a_0$ is just the volume of $X$ and 
$$
a_1=\frac{1}{2\pi}\int_X S(\omega),
$$
where $S(\omega)$ is the scalar curvature of $\omega$. We will use the following result (Prop. 6 in \cite{Donaldson1}):
\begin{prop}
\begin{enumerate}
  \item $\rho(\omega)$ has an asymptotic expansion as $k \to \infty$
  $$
   \rho_k(\omega) \sim A_0(\omega)k^n +A_1(\omega)k^{n-1}+\cdots
  $$
  where $A_i(\omega)$ are smooth functions on $X$ defined locally by $\omega$. In particular, 
  $$
  A_0(\omega)=1,\qquad A_1(\omega)=\frac{1}{2\pi}S(\omega).
  $$
  \item The expansion holds uniformly in the $C^\infty$ norm; in that for any $r,N>0$
  $$
  \left\| \rho_k(\omega)-\sum_{i=0}^N A_i(\omega)k^{n-i} \right\|_{C^r(X)}\leqslant K_{r,N,\omega}k^{n-N-1}
  $$
  for some constants $K_{r,N,\omega}$. 
\end{enumerate}
\end{prop}

Now assume that we are given balanced metrics $\omega_k$ converging to $\omega_\infty$. Then by the previous proposition
$$
\left\| \rho_k(\omega_k)-k^{n}-\frac{1}{2\pi}S(\omega_k)k^{n-1} \right\|_{C^0(X)}\leqslant ck^{n-2}
$$
for some constant $c$. Since $\omega_k$ is balanced $\rho_k(\omega_k)$ is constant: $\rho_k(\omega_k)= \dim H^0(X,L^k)/V$, and we can use (\ref{eq-rho}) to find that
$$
\left\|  \ff 1 V (V k^n + a_1k^{n-1}+\cdots )-k^{n}-\frac{1}{2\pi}S(\omega_k)k^{n-1} \right\|_{C^0(X)}\leqslant ck^{n-2},
$$
or equivalently
$$
\left\|  \ff {2\pi} V  a_1-S(\omega_k) \right\|_{C^0(X)}= O(k^{-1})
$$
Hence $S(\omega_\infty)=S_0=constant$, where $S_0=\frac{1}{V}\int_X S(\omega)$ is the mean curvature.

\section{The hermitian Yang-Mills equations}

We are now ready to generalize the T-operator, which provided an
approximation scheme for the constant curvature metric, to a
``generalized T-operator'' which can be used to find a solution of the
Yang-Mills equations on a Calabi-Yau manifold $X$.

We briefly recall the argument that a solution of the Yang-Mills equations
which preserves ${\cal N}=1$ supersymmetry, must be hermitian Yang-Mills.
First, the supersymmetry variation of the gaugino has to vanish,
$$
\Gamma^{\mu\nu} F^a_{\mu\nu}\epsilon =0,
$$
where $F^a_{\mu\nu}$ is the Yang-Mills field strength, and
$\epsilon$ is the covariantly constant spinor.

Going to complex coordinates $(i,\bi)$ and 
rewriting of the Clifford algebra as
$$
\Gamma_i \rightarrow dz^i; \qquad
 \Gamma_{\bi} \rightarrow \omega_{\bi j} ~ {\partial}^j ,
$$
this is equivalent to
$$
F_{ij} = F_{\bi\bj} = 0; \qquad
\omega^{i\bj} F_{i\bj} = 0 .
$$
This is the particular case of the hermitian Yang-Mills equations
with $\Tr F = 0$.  The general case replaces the last equation with
$$
\omega^{i\bj} F_{i\bj} = c\cdot{\bf 1} 
$$
for a constant $c$, determined by the first Chern class.  For convenience
we abbreviate this equation below as
$$
\bigwedge F = c\cdot{\bf 1} .
$$

Next we review the relation between solutions of these equations,
and holomorphic bundles carrying hermitian-Einstein metrics.
In physics, one defines Yang-Mills theory in terms of a connection
on a vector bundle with a fixed metric.
First, a connection on a vector bundle can be described in terms
of a connection one-form by choosing a frame for the
bundle, say $e_a(x)$, and defining the covariant derivative as
$$
D(v^a e_a) 
 = (dv^a) e_a + v^a A^b_a e_b .
$$
In physics, one usually takes the frame to be orthonormal, 
$(e_a,e_b) = \delta_{ab} $, and thus
\be\label{eq:physmet}
(u,v) = (u^a e_a,v^b e_b) = (u^a)^* v^a ,
\ee
where $*$ is complex conjugation.

The condition that the connection be compatible with the metric,
\be\label{eq:metcomp}
d(u,v) = (Du,v) + (u,Dv),
\ee
reduces to requiring the connection one-form to be anti-hermitian,
\be\label{eq:antiherm}
\Aphys_i = -\Aphys_i^\dag.
\ee

In mathematics, one often considers a more general frame, for
which the metric is a hermitian matrix,
\be\label{eq:genherm}
(e_a,e_b) = G_{\ba b}, \qquad G = G^\dag .
\ee
Decomposing the positive definite hermitian matrix $G$ as
\be\label{eq:Ghh}
G = h^\dag h,  
\ee
we see that the math and physics conventions differ
by a complex gauge transformation:
$ 
u = h\,s.
$ 
This complex gauge transformation  leads to a different form
for the connection, according to the standard
relation
\be\label{eq:compgy}
\partial_i + \Amath_i = h (\partial_i + \Aphys_i) h^{-1} .
\ee

Now, equations of the form
$$
F_{\bi\bj} = 0\  \forall\ \bi,\bj
$$
will be integrability conditions for the covariant derivatives.
In particular, this equation has the general solution
$$
\partial_{\bi} + \Aphys_{\bi} = g^{-1} \bar\partial_{\bi} g ,
$$
in other words the $\bD$ covariant derivatives are obtained from
the derivative $\bar\partial$ by a complex gauge transformation.

Thus, we can use \eq{compgy} to bring the connection to the gauge
$\bAmath=0$, at the cost of losing the simple metric \eq{physmet} and
\eq{antiherm}.  Actually, the covariant
derivative is still compatible with the metric as in \eq{metcomp},
we just have a non-trivial fiber metric $G$.
The metric compatibility condition becomes
$$
0 = \partial(u,v) = ({\bpartial u},v)+(u,D v)
$$
so
$$
\partial G_{\ba b} = G_{\ba c} \Amath^c_b
$$
or equivalently
$$
\Amath = G^{-1}\partial G .
$$

Conversely, if we are given a metric $G$, then we can use
the inverse complex gauge transformation to bring the connection
back to the unitary form.  
This leads to the formula
$$
\bAphys_{\bi} = h (\bpartial_{\bi} h^{-1}).
$$
Using \eq{antiherm}, we can get the entire connection, so the metric
contains the same information as a connection satisfying 
$F^{(0,2)}=F^{(2,0)}=0$.  Thus we can rephrase the final equation on
$F^{(1,1)}$, as a condition on the metric.  It is simplest to write
this in the ``mathematical'' gauge $\bAmath=0$, in which it is
\be \label{e5}
c\cdot{\bf 1} = \omega^{i\bj}F_{i\bj}
 =\omega^{\bj i}\bpartial_{\bj} \Amath_i
 =\omega^{\bj i}\bpartial_{\bj} \left(G^{-1}\partial_i G \right) .
\ee
A metric $G$ satisfying this equation is a ``hermitian-Einstein'' 
metric.  It is simply related to a hermitian Yang-Mills connection
as above.

Finally, using the complex gauge transformation above, the standard
physical inner product
\be\label{eq:physvip}
\vev{u|v} \equiv \int_X (u^*)_a v^a
\ee
is equal to the natural inner product generalizing \eq{defH1},
\begin{eqnarray}\label{eq:mathvip}
\vev{u|v} &=& \vev{h~ s|h~ t} \\
 &=& i\int_X G_{a\bb}~ \bs^\bb~ t^a \,d{\vol_X}.
\end{eqnarray}

\subsection{Embedding vector bundles}

We now want to represent the hermitian metric $G_{a\bb}$ in 
the same way as we did for line bundles, by 
introducing a complete basis of sections.
Now an irreducible bundle $E$ with $c_1=0$, and thus of interest 
for string compactification, will not have global sections.
What we do instead is to make the same construction for 
$E(k) \equiv E \otimes \cL^k$, which will have global sections.
We can again think of these sections as a basis of polynomials 
approximating functions on which to base our numerical scheme.

Thus, consider a rank $r$ vector bundle $E$, and suppose 
that $E(k)$ has $N$ global sections.
Choosing a local frame as above,
a basis for these will be an $N$ by $r$ matrix $z_\alpha^a$.
This is defined up to a $GL(N)$ change of basis, and up to a 
$GL(r)$ change of frame. After making these identifications, such a
matrix $z$ defines a point in the Grassmannian $G(r, N )$  of $r$ planes in $\C^N$.

Given a metric $G_{a\bb}$ on the fibers of $E(k)$,
we can define the matrix of inner products
$$
H_{\alpha\bbeta} = \vev{z_\beta|z_\alpha}
$$
as above.  Such a metric could be obtained by multiplying a metric
$G^{(0)}$ on $E$, by one on $\cL^k$ as defined earlier.  Or, it
might simply be an $r\times r$ hermitian matrix of functions (in
each local frame) with appropriate transformation properties.

Now there is a natural set of metrics on $E(k)$ generalizing
\eq{Ks}, again parameterized by an $N\times N$ matrix,
defined by
$$
(G^{-1})^{a\bb} = g^{\alpha\bbeta} z_\alpha^a (z^\dag)_\bbeta^\bb\, ,
$$
where the dagger is hermitian conjugation.  Again, the approach will
be to find a natural metric in this class which is a good approximation
to the hermitian-Einstein metric. This will lead to a hermitian Yang-Mills
connection on $E(k)$.  But this is simply related to the hermitian
Yang-Mills connection on $E$, because twisting by $\cL^k$ only
modifies the trace part of the field strength.

\subsection{Generalized T-operator}

We will now turn to a proposal for a generalized
T-operator, which produces the hermitian-Einstein metric on a stable
vector bundle. To begin with we use results by Wang about balanced
metrics on such bundles \cite{Wang:metricsbundles}.

We consider again a polarized $n$ dimensional manifold $(X,\cL)$ and an
irreducible holomorphic vector bundle $E$ of rank $r$ on $X$. Then by
Kodaira embedding we know that for k sufficiently large, a basis
$z_{\alpha}^a$ of the global sections of $E(k)$ will
give rise to an embedding
$$
\xymatrix{
        X~       \ar@{^{(}->}[r]^-i       &     G(r, N ) .   \\
}
$$
Now Wang proves the following:
\begin{theo}\label{bundl1}
$E$ is Gieseker stable iff there is an integer $k_0$ such 
that for $k > k_0$, the {\em k}th embedding given as above can be moved to a balanced 
place, i.e.,  there is a $g \in SL(N ,\C)$ which is unique up to left translation by $SU(N)$ such that: 
$$ \frac{1}{V} \int_{g\cdot X} {z (z^\dag z)^{-1} z^\dag\ d V }=
\frac{r}{N} I_{N \times N} .
$$ 
\end{theo} 

We call the equation above the ``balance equation.''
In the case that $E$ is a line bundle,
this definition reduces to that of a 
balanced  embedding in $\mathbb{P}^{N-1}$. 

Now, let $h$ be a hermitian  metric on $\cL$ 
and $H$ be a hermitian  metric on $E $, and fix the K\"ahler 
form on $X $ to be $\omega = \ff i {2\pi} Ric(h)$. 
Let $\vol$ denote the volume of $(X, \omega)$. 
Suppose 
${S_1 , \ldots , S_N }$ is an orthonormal basis of $H^0(X,E (k))$  with respect to the induced $L^2$ -metric $\langle . \, ,  . \rangle$.
The Szeg\"o kernel $B_k$ is a generalization of the function $\rho(\omega)$ defined in \eq{bk}. It is defined as the fiberwise homomorphism
$$
B_k(x) = \sum_{i=1}^N \langle . , S_i(x)\rangle S_i (x) \colon E_x \to E_x .
$$ 
This expression is independent of the choice of orthonormal basis.

Now the local 
form of Theorem \ref{bundl1}  can be stated as follows (Corollary 1.1 of \cite{Wang:metricsbundles}):
\begin{theo}
E is Gieseker stable iff there is an integer $k_0$ such that for any
$k > k_0$, we can find a metric $H^{ (k)}$, which we will call the {\em
balanced metric} on $E (k)$, such that the Szeg\"o kernel satisfies
the equation
$$
B_k (x)  = \frac{\chi(k)}{V~ r} I_E
$$ 
where $I _E$ is the identity bundle morphism 
and $\chi(k)$ is the Hilbert polynomial of 
$E$ with respect to the polarization $\cL$. 
\end{theo}

The theorem tells us that if $E$ is Gieseker stable then for large $k$
there is a balanced metric $H^{ (k)}$ on $E (k)$. Hence we will have a
sequence of hermitian metrics $H_k := H^{ (k)}\otimes h^{-k} $ on $E$.
The importance of the balanced metric $H^{ (k)}$ for physical
applications follows from the following theorem:
\begin{theo}
Suppose $E$ is Gieseker stable. If $H_k \to H_\infty$ in the
$C^\infty$ norm as $k \to \infty$, then the metric $H_\infty$ solves
the ``weak hermitian-Einstein equation'',
\begin{equation}
\label{hym}
\frac{i}{2\pi} \bigwedge F_{(E,H_\infty)} + \frac{1}{2}S(\omega) I_E=
\left( \frac{deg(E)}{V r}+\frac{\bar{s}}{2} \right) \cdot I_E
\end{equation}
where $\bigwedge F_{(E ,H_\infty)}$ is the contraction of the
curvature form of E with $\omega$, $S (\omega)$ is the scalar
curvature of $X$ and $\bar{s} := \frac{1}{V} \int_X S
\frac{\omega^n}{n!}$ .  Conversely, suppose there is a hermitian
metric $H_\infty$ solving this equation, then $H_k \to H_{\infty}$ in
$C^r$ norm for any $r$.
\end{theo}

To prove (\ref{hym}) one can work along the same lines as in the proof
of Theorem~\ref{theo2}, using Catlin's and Wang's results for the
expansion of the Szeg\"o kernel.
\begin{prop}

\begin{enumerate}
\item For fixed hermitian metrics $H$ and $h$ on $E$ and ${\cal
O}_X(1)$ respectively, there is an asymptotic expansion as $k \to \infty$
  $$
  B_k(H,h)\sim A_0(H,h)k^n +A_1(H,h)k^{n-1}+ \cdots,
  $$
  where $A_i(H,h) \in \Gamma(\End E)$ are smooth sections defined locally by $H$. 
In particular,
  $$
  A_0(H,h)=I_E,\;\;A_1(H,h)=\frac{i}{2\pi}\bigwedge F(E,Ric(h)) +\frac{1}{2}S(\omega) \cdot I_E
  $$
\item The expansion holds uniformly in the $C^\infty$ norm; in the sense that for any $r,N>0$
  $$
  \| B_k(H,h) - \sum_{i=0}^N A_i(H,h)k^{n-i}      \|_{C^r}\leqslant K_{r,N,H,h} k^{n-N-1}
  $$
  for some constants $K_{r,N,H,h}$. 
\end{enumerate}
\end{prop}

Now we can repeat the steps of the argument outlined in Section~\ref{s:prs}. Under the assumption that $H_k \to H_\infty$ in $C^\infty$ we find that for  $r>0$ 
$$
\| B_k(H_k) - I_E k^{n}-  \frac{i}{2\pi}\bigwedge F(E,Ric(h)) +\frac{1}{2}S(\omega) \cdot I_E  k^{n-1}  \|_{C^r}\leqslant C k^{n-2}
$$
for some fixed constant $C$. By assumption $H^{(k)}$ is balanced, hence $B_k(H_k)=\chi(k)/r V I_E$. This implies that 
$$
\|\frac{i}{2\pi} \bigwedge F_{(E,H_\infty)} + \frac{1}{2}S(\omega) I_E-
\left( \frac{deg(E)}{V r}+\frac{\bar{s}}{2} \right) \cdot I_E\|=O(k^1).
$$

\subsubsection{Generalized T-operator }

Using the strong analogy between the construction of metrics with
constant Kahler curvature and metrics on stable bundles which obey the
hermitian-Einstein equation, we propose the following generalized
T-operator:
\begin{equation}\label{e4}
 T(G)=\frac{N}{V r}\int_X {z (z^\dag G^{-1} z)^{-1} z^\dag \ d V },
\end{equation}
where as before, $z$ is an $N$ by $r$ matrix of holomorphic sections
of $E$.

The relevance of this proposal follows from the following conjecture:
\begin{conj}\label{gT}
If a balanced embedding $i\colon X \hookrightarrow G(r,N)$ exists,
then for every hermitian $N \times N$ matrix $G$, the sequence
$T^r(G)$ converges to a fixed point $G_0$ as $r\to\infty$. Using an
orthonormal basis with respect to $G_0$, the embedding is balanced,
and as outlined above, it provides an approximate solution for the
corresponding hermitian-Einstein equation.
\end{conj}
This conjecture may require additional technical assumptions, such
as the earlier one of $\Aut(X,E)$
being discrete.  We have not attempted to prove it, but would hope
that this can be done along the same lines as \cite{Donaldson1,Sano}.

In the following section we will numerically test the conjecture for
several stable vector bundles on $\mathbb{P}^2$, and on the Fermat
quintic in $\P^4$, and find that it works for these cases.

\section{Examples}

\subsection{Hermite-Einstein metric on the tangent bundle of $\mathbb{P}^n$}

Let $\mathbb{P}^n$ be the complex projective space of dimension $n$, and $\{ Z_i \}_{i=0}^{i=n}$ 
its homogeneous coordinates. We will work on the open set $Z_0 \neq 0$  and chose the  local inhomogeneous coordinates
$w_{i}=Z_{i}/Z_{0}$. The Fubini-Study metric on $\P^n$  
$$
g_{i\bar{j}} = \frac{1}{1+\sum_i |w_i|^2} \delta_{i\bar{j}} - \frac{w_{i}\bar{w}_{j}}{(1+\sum_i |w_i|^2)^2}.
$$
is the unique maximally symmetric metric, with its group of Killing symmetries  isomorphic to $U(n+1)$. In addition, this metric is Einstein, that is its Ricci tensor is proportional to the metric itself. Therefore its associated curvature tensor  
obeys the hermitian Yang-Mills  equation. The Donaldson-Uhlenbeck-Yau theorem then implies that the tangent bundle of $\P^n$, $T\mathbb{P}^n$, is a rank $n$ stable bundle on $\mathbb{P}^n$.\footnote{
The stability of $T\mathbb{P}^n$ also has purely algebraic proof.}
It follow from this that  the balanced metric on the bundle $T\mathbb{P}^n$ must be the Fubini-Study metric.

To describe the tangent bundle  $T\mathbb{P}^n$ we  use the Euler sequence 
\begin{equation}\label{eulerseq}
0 \longrightarrow \mathcal{O}_{\mathbb{P}^n} \longrightarrow \mathcal{O}_{\mathbb{P}^n}(1)^{\oplus (n+1)}
\longrightarrow T\mathbb{P}^n \longrightarrow 0.
\end{equation}
Here $ \mathcal{O}_{\mathbb{P}^n}(1)$ denotes the hyperplane line bundle. After twisting the sequence
by $\mathcal{O}_{\mathbb{P}^n}(k)$ and taking the cohomology we find the short exact sequence (SES)
\begin{equation}\label{e1}
0 \longrightarrow H^0({\mathbb{P}^n},\mathcal{O}_{\mathbb{P}^n}(k)) \longrightarrow H^0({\mathbb{P}^n}, \mathcal{O}_{\mathbb{P}^n}(k+1)^{\oplus (n+1)})
\longrightarrow H^0({\mathbb{P}^n},T\mathbb{P}^n(k)) \longrightarrow 0.
\end{equation}
This gives an explicit description for $H^{0}(\P^n,T\mathbb{P}^n(k))$, which for sufficiently large $k$ gives the embedding
\begin{equation}\label{e2}
\mathbb{P}^n \hookrightarrow \mathrm{G}(n,\, W)
\end{equation}
where $W=H^{0}({\mathbb{P}^n},T\mathbb{P}^n(k))^*$, and $\mathrm{G}(n,\, W)$ is the Grassmannian of $n$-planes in $W$. 

Based on (\ref{e1}), we choose to describe the global sections of 
$T\mathbb{P}^n(k)$ by an $n+1$ vector 
$$
(M_0,\ldots,M_n)
$$ 
where $\{M_i\}_{i=1}^n$ are arbitrary monomials of degree $k+1$ in
the homogeneous coordinates $Z_i$, while $M_0$ is any degree $k+1$
monomial which does not contain an $Z_0$.

Now we show  how to construct the embedding (\ref{e1}) for any $k\geq 0$. We start by choosing a frame $\{ \hat{e}_i \}_{i=0}^{n}$ for the  vector bundle 
$\mathcal{O}(k+1)^{\oplus (n+1)}$. This amounts to choosing a section for every one the $n+1$ summands. For simplicity we chose the same section in every  summand. The Euler sequence (\ref{eulerseq}) imposes the condition 
$$ 
\sum_{i=0}^n Z_i \hat{e}_i =0.
$$
Locally this gives a frame for $T\mathbb{P}^n$, if we solve for
$$
\hat{e}_0 = -\sum_{i=1}^n \frac{Z_i}{Z_0}\hat{e}_i = -\sum_{i=1}^n w_i \hat{e}_i.
$$
Expanding the global sections of $T\mathbb{P}^n(k)$ in the local frame $\{\hat{e}_i \}_{i=1}^n$ gives an $n \times \dim(W)$ matrix, which is the explicit realization of our embedding \cite{GH}. 

To illustrate the procedure consider $T\mathbb{P}^2(0)$. $\mathcal{O}_{\P^2}(1)$ has 3 global sections: $Z_0,Z_1,Z_2$. Choosing $Z_0$ to be the local frame in every summand of $\mathcal{O}_{\P^2}(1)^{\oplus 3}$, and discarding the global section $Z_0$ from the first $\mathcal{O}_{\P^2}(1)$, we find the matrix
\begin{equation}
z=\left(\begin{array}{cccccccc}-w_1^2 & -w_1w_2 & 1 & w_1 & w_2 & 0 & 0 & 0 \\-w_1w_2 & -w_2^2 & 0 & 0 & 0 & 1 & w_1 & w_2\end{array}\right)
\end{equation}

For an initial hermitian metric $G_{0}$ on the vector space 
$W=H^{0}(\P^n, T\mathbb{P}^n(k))^*$, our generalized T-operator  (\ref{e4}) gives the iterations 
$$
G_{m+1}=T(G_m)=
\frac{ \dim W}{n\, \mathrm{Vol}(\mathbb{P}^n)}\int_{\mathbb{P}^n} {z (z^\dag G_m^{-1} z)^{-1} z^\dag \ d V }.
$$
 
We tested the converges of the $T$-map starting with $G_{0}=I$ in the case $n=2$ for $k=1,\ldots ,5$. In all cases we converged to a given $G_{\infty}$ for less than 10 iterations, with a precision of 0.1\%. 

The balanced metric $H^{(k)}$ on the vector bundle $T\mathbb{P}^n(k) $ induced by $G_{\infty}$ is given by
\begin{equation}
H^{(k)}=(z^\dag G_{\infty}^{-1} z)^{-1} .
\end{equation}
Let $h$ be Fubini-Study metric on the hyperplane bundle $\mathcal{O}_{\P^n}(1)$, that is the metric with constant scalar curvature. 
Then the metric
$$
H_k := H^{(k)}\otimes h^{-k}=(z^\dag G_{\infty}^{-1} z)^{-1}\otimes h^{-k}
$$
is the balanced metric on $T\mathbb{P}^n$. Our numerical computations show that this is indeed  the Fubini-Study metric  on $T\mathbb{P}^n$, as explained earlier. The numerical  agreement is within 0.5\%. This provides the first non-trivial test of our conjecture.

\subsection{A stable rank 3 bundle over $\mathbb{P}^2$}

In this section we test our generalized T-operator on a rank 3 vector
bundle $V^*$ over $\mathbb{P}^2$.  We first consider its dual $V$, defined by
four linearly independent global sections $\{m_i\}$ of 
${\cal O}_{\mathbb{P}^2}(2)$ through the SES
$$
\xymatrix{
0\ar[r] &V \ar[r] &{\cal O}_{\mathbb{P}^2}^{\oplus 4} \ar[r]^-m &{\cal O}_{\mathbb{P}^2}(2) \ar[r] & 0.
}
$$
This bundle has moduli, which are implicitly determined by the choice
of the sections $\{m_i\}$.  Before choosing these, let us
check stability, which does not depend on the specifics of this choice.

To check stability, we have to ensure that neither $V$ nor $\wedge^2V$
have destabilizing line bundles. Using the canonical isomorphism
$$
\wedge^2 V = \det V \otimes V^*
$$
we find  the slopes
$$
\mu(V)=-2/3,\,\,\,\mu(\wedge^2 V)=-4/3.
$$
Since $Pic(\mathbb{P}^2)=\mathbb{Z}$,  all line bundles are of the form ${\cal O}_{\mathbb{P}^2}(p)$ for some $p$.
Hence it is sufficient to show that 
$$
H^0({\mathbb{P}^2},V)=0,\;\;\;H^0({\mathbb{P}^2},\wedge^2 V(1))=0.
$$
The first fact follows from the defining sequence of $V$, if we assume that $\{m_i\}$ are linearly independent. To prove the second statement we use 
$$
H^0({\mathbb{P}^2},\wedge^2 V(1))=H^0({\mathbb{P}^2},V^*(-1))=H^2({\mathbb{P}^2},V(-2))^*.
$$
Again, this statement follows easily from the defining sequence of $V$.
Finally, stability of $V$ implies stability for $V^*$. 

We will now compute the hermitian Yang-Mills  connection on $V^*$ using our generalized T-operator. 
First observe that $V^*(k)$ fits into the short exact sequence
\begin{equation}\label{Vdual}
0\to {\cal O}_{\mathbb{P}^2}(k-2) \to {\cal O}_{\mathbb{P}^2}(k)^{\oplus 4} \to V^*(k)\to 0.
\end{equation}
This leads to another SES
$$0\to 
H^0({\mathbb{P}^2},{\cal O}_{\mathbb{P}^2}(k-2)) \to H^0({\mathbb{P}^2},{\cal O}_{\mathbb{P}^2}(k)^{\oplus 4}) \to H^0({\mathbb{P}^2},V^*(k))\to 0.
$$
We can use this expression for an explicit parameterization of $H^0({\mathbb{P}^2},V^*(k))$.  

For concreteness let us choose to be  four global sections $\{m_i\}_{i=1}^4$ defining $V$  to be 
$$
Z_1Z_2,\;Z_0Z_1,\; Z_0Z_2,\;Z_0^2.
$$
Now we choose a frame $\{\hat{e_i}\}$ for ${\cal O}_{\mathbb{P}^2}(k)^{\oplus 4}$. The defining equation (\ref{Vdual}) of $V^*(k)$ imposes the condition $\sum_i m_i \hat{e_i} =0$, and gives a frame for $V^*$. Locally we can solve for $\hat{e}_0$, and working in inhomogeneous coordinates $w_i$ we find that
$$
\hat{e}_0= -\frac{1}{w_2}\hat{e}_1-\frac{1}{w_1}\hat{e}_2-\frac{1}{w_1w_2}\hat{e}_3.
$$
Expanding the global sections of $H^0({\mathbb{P}^2},V^*(k))$ in the frame $\{\hat{e}_i\}_{i=1}^3$ gives a matrix, which is the embedding map. 

We studied the convergence of our generalized $T$-operator numerically for $k=2,3$ and 4, and found that convergence was achieved for less than 10 iterations. As before, the  metric on $V^*(k)$ is given by
\begin{equation}
H^{(k)}=(z^\dag G_{\infty}^{-1} z)^{-1} ,
\end{equation}
while the corresponding metric on $V^*$ is
$$
H_k := H^{(k)}\otimes h^{-k}=(z^\dag G_{\infty}^{-1} z)^{-1}\otimes h^{-k},
$$
where $h$ is again the Fubini-Study metric on ${\cal O}_{\mathbb{P}^2}(1)$.

Since in this case the balanced metric on $V^*$ is not a priori known,
one needs a different approach, than used in the previous section for
$T\P^2$, to test how close is the approximate balanced metric to
satisfying the hermitian Yang-Mills equation. But this quite easy to
do numerically once the balanced metric $G_{\infty}$ is known, as all
we need to do is to check how close we are to satisfying
Eq.~(\ref{e5}). In all cases considered Eq.~(\ref{e5}) was satisfied
to within 1\% accuracy.

\subsection{A rank 3 bundle on the Fermat quintic}

In this section we turn to a much more complicated case than our previous examples, that of a stable rank 3 bundle on the Fermat quintic $Q$ in $\P^4$:
\begin{equation}
Q:\qquad  Z_0^5+Z_1^5+\cdots +Z_4^5=0.
\end{equation}

Testing our generalized T-operator in this case necessitates knowledge of the Ricci flat metric on the Fermat quintic. For this we use Donaldson's original T-operator  \eq{defH}, which we turn to first.

\subsubsection{Ricci flat metric on Fermat quintic}

We consider the embedding of $Q$ given by the complete linear system of cubics, $H^{0}(Q, \mathcal{O}_Q (3))$, whose
complex projectivization is isomorphic to $\mathbb{P}^{34}$. The balanced metric will be the restriction of a Fubini-Study metric on $\mathbb{P}^{34}$. An indirect test that this has indeed vanishing Ricci curvature is included in the next section. 

In order to do practical calculations with  Donaldson's  T-map, we have to perform the  integrals on $Q$ numerically. We introduce
a discrete approximation to the Calabi-Yau volume form $d\mu_{\Omega}=\Omega\wedge\bar{\Omega}$,
defining it by a weighted set of $M$ points $\{ x_a \}_{a=1}^M\in Q$, with masses $\nu_a$:
\begin{equation}\label{numint}
\int_Q \, \big( \,\,\, \big) \, d\mu_{\Omega} \approx \sum_{a=1}^M\, \big( \,\,\, \big) \, \delta(x-x_a)\nu_a.
\end{equation}
This numerical measure  gives an accurate approximation to the analytical one for large $M$.
In our computations we build $10$ different samples of 100,000 points, which we use independently to 
iterate the T-map until convergence is reached, i.e.,  the sequence $\{ T^{r}(G_0) \}_{r=0}$
obeys
$$
\vert\vert T^{r+1}(G_0) - T^r (G_0) \vert \vert < \epsilon.
$$
In our simulations the fixed point of this discrete version of the $T$-map was  
reached after 15-20 iterations.
Each weighted point set gave rise to a convergent sequence. 
The 10 different hermitian forms $\{ G^{e}_{\infty} \}_{e=1}^{10}$ 
approximating the balanced metric in $\P H^{0}(Q,\mathcal{O}_Q (3))$ agree up to 
\begin{equation}
\label{errordef}
\mathrm{max}\, \Bigg[ 
\frac{\sigma(G^{e}_{\infty})}{\vert \langle G^{e}_{\infty} \rangle \vert} \Bigg]
\approx 0.9 \% ,
\end{equation}
where $\vert \langle G^{e}_{\infty} \rangle \vert$ is the average matrix of the 
ten different outputs and $\sigma(G^{e}_{\infty})$ is the standard deviation matrix.
The ratio $\sigma(G^{e}_{\infty})/ \vert \langle G^{e}_{\infty} \rangle \vert$ is computed entry by entry,
and the maximum is taken over all entries.
We used the average $\langle G^{e}_{\infty} \rangle$ as approximation for the balanced metric on
$H^{0}(Q,\mathcal{O}_Q (3))$.

\begin{figure}[h]
      \centering
      \includegraphics[width=\textwidth]{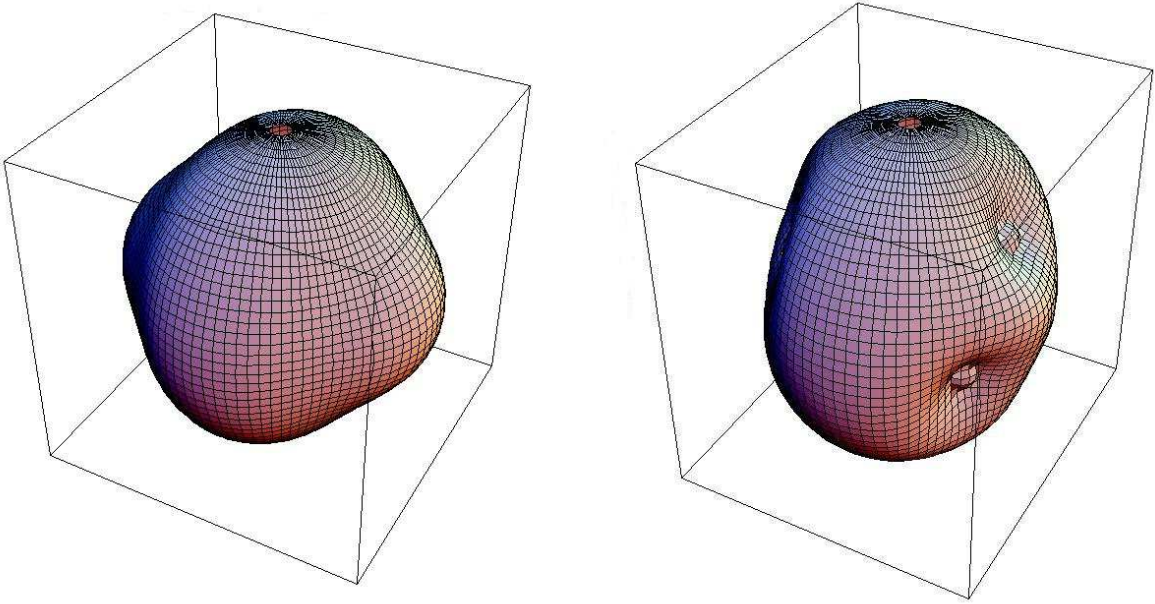}
      \caption{The shape of the rational curve for the balanced and non-balanced metrics.} \label{p:bm}
\end{figure}

To get a visual picture of the geometry implicit in the construction, 
we consider the rational curve $t\colon\mathbb{P}^1 \hookrightarrow Q$,
defined locally by the parametrization 
\begin{equation}
\label{RatCur}
\left(\begin{array}{ccccc} 1, & -1, &  t, & 0, & -t
\end{array}\right)
\end{equation}
with $t\in \mathbb{C}\cup\infty$. Take the sections $Z_1^3+Z_4^3$ and
$Z_0^3$ from $H^{0}(Q,\mathcal{O}_{Q}(3))$, and consider the function
$s=(Z_1^3+Z_4^3)/Z_0^3=w_1^3+w_4^3 $. In Fig.~\ref{p:bm} we consider
the real function $|s|^{2}_{G}$ restricted to the rational curve
(\ref{RatCur}), where we take the stereographic projection of the
complex $t$-plane and for a given $t$ we plot $|s(t)|^{2}_{G}$ in the
radial direction. For the balanced metric $\langle G^{e}_{\infty}
\rangle$ the deviation from being spherical is small. For comparison
we also show the same plot for the case of a generic non-balanced
hermitian form $G$ with random entries.

\subsubsection{Solution of the hermitian  Yang-Mills  equation}

In this section we use the generalized T-operator to produce a hermitian  Yang-Mills  connection on a rank three stable vector bundle $V$ on the Fermat quintic $Q$. We also implicitly test that the previously obtained balanced metric on $Q$ indeed has vanishing Ricci curvature.

We define the rank three bundle $V$  by the following SES
$$
\xymatrix{
0\ar[r] &{\cal O}_Q(-1) \ar[r]^-{\beta} & {\cal O}^{\oplus 4}_Q  \ar[r] & V \ar[r] & 0.
}
$$
$\beta$ is given by four generic global sections of ${\cal O}_Q(1)$, which do not intersect on $Q$, hence $V$ is indeed a vector bundle. In addition, the  first Chern class of $V$ is $
c_1(V)=H,
$
hence $V$ is not a simple twist of the tangent bundle of $Q$.
That fact that $V$ is stable was proved in \cite{DRY}.

Once again, we use 
$$
\xymatrix{
0\ar[r] &{\cal O}_Q(k-1) \ar[r] & {\cal O}^{\oplus 4}_Q(k)  \ar[r] & V(k) \ar[r]& 0		,}
$$
and its associated long exact sequence in cohomology 
$$
\xymatrix{
0\ar[r] &H^0(Q,{\cal O}_Q(k-1)) \ar[r] & H^0(Q,{\cal O}^{\oplus 4}_Q(k))  \ar[r] & H^0(Q,V(k))\ar[r]& 0,
}
$$
to derive a frame for $V$ and an explicit parameterization for the global sections.
We choose $\beta=(Z_0,\ldots, Z_3)$. Using the frame $\{\hat{e}_i\}_{i=0}^4$ for ${\cal O}^{\oplus 4}_Q$, we also get a frame for $V$ with the relation
$$
\hat{e}_0=-\sum_{i=1}^3 w_i \hat{e}_i.
$$

In this paper we restrict to the case $k=1$ for which $\dim H^{0}(Q,V(1))=19$. The coordinate matrix 
\begin{equation}
\label{embbedG }
z(w)=\left(\begin{array}{ccccccc}1 \ldots w_4 & 0 & 0 & -w_1^2 & -w_1w_2 & -w_1w_3 & -w_1w_4 \\0 & 1 \ldots w_4 & 0 & -w_1w_2 & -w_2^2 & -w_3w_2 & -w_4w_2 \\0 & 0 & 1 \ldots w_4 & -w_1w_3 & -w_2w_3 & -w_3^2 & -w_4w_3\end{array}\right)
\end{equation}
gives the embedding into the Grassmannian $Q\hookrightarrow G(3,19)$.

Using the integration techniques described in the previous section,  we iterate the generalized T-operator. We reach the fixed point after 12-15 iterations for several different samples of weighted points which approximate
the analytical measure, allowing us to estimate the balanced metric for $H^{0}(Q,V(1))$ with an error
of 1.1\%.\footnote{We estimate the errors using (\ref{errordef}).} 

The metric on $V(1)$ is given by
\begin{equation}\label{HEmet}
H=(z^\dag G_{\infty}^{-1} z)^{-1} ,
\end{equation}
To test the accuracy of this metric  we  evaluate the right hand side of the hermitian  Yang-Mills equations (\ref{e5}).
We find  the mean to be
$$
\langle \omega^{i\bj}F_{i\bj} \rangle=\frac{1}{\mathrm{Vol}(Q)}\int_Q \big( \omega^{i\bj}F_{i\bj} \big) d\mu_{\Omega} \approx 1.31\times \mathbf{I}_{3\times 3}
$$
with $\mathbf{I}_{3\times 3}$ the $3\times 3$ identity matrix. In our conventions the theoretical value of the constant is 4/3. The standard deviation of the individual matrix elements is
$$
\sigma\big( \omega^{i\bj}F_{i\bj} \big) =\mathrm{max}\left[\sqrt{\frac{1}{\mathrm{Vol}(Q)}\int_Q \Big( \omega^{i\bj}F_{i\bj} - \langle \omega^{i\bj}F_{i\bj} \rangle \Big)^2 d\mu_{\Omega}}\ \right] \approx 0.15,
$$
where the  square-root and the square are performed entry by entry.
Therefore, $\omega^{i\bj}F_{i\bj}$ is a global constant on $Q$ times the identity, within an error of  $ 0.15/1.31\approx 11 \%$. This implies that the hermitian Yang-Mills equation (\ref{e5}) is satisfied with this  accuracy. 

Testing the hermitian Yang-Mills equation provides an implicit test of Ricci flatness, since it is precisely the Ricci flat metric that is needed in the hermitian Yang-Mills equation. If we had gotten this metric wrong, then we would have had no chance of satisfying the hermitian Yang-Mills equation.

\begin{figure}[h]
      \centering
      \includegraphics[width=0.35\textwidth, angle=105]{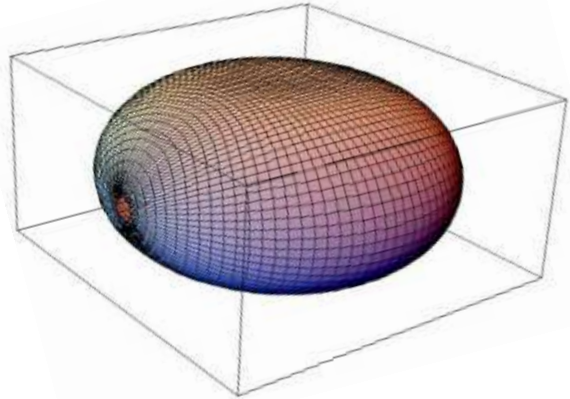}
      \caption{The probability density 
                    in the radial direction on the rational curve.}\label{p:wf}
\end{figure}

Finally, to visualize the construction,  
in Fig.~\ref{p:wf}  we took the rational curve defined in (\ref{RatCur}), and we plotted
the function $|\Psi|_G^2$, where
$$
\Psi =(w_1 w_4)\hat{e}_1 + (w_2 w_4)\hat{e}_2 + (w_3 w_4)\hat{e}_3 
$$
and $G$ is the balanced metric we obtained.
If we interpret $\Psi$ as a wave-function, then Fig.~\ref{p:wf} exhibits the probability density $\langle \Psi \vert \Psi \rangle $ in the radial direction, restricted to the rational curve (\ref{RatCur}).

\subsubsection*{Acknowledgements}

We would like to thank B. Shiffman and S. Zelditch for valuable
advice at the beginning of this project.
This research was supported in part by the DOE grant DE-FG02-96ER40949.


\end{document}